\documentclass[12pt,eqsecnum,floats,aps,amsmath,amssymb,nofootinbib,prd]{article}

\usepackage{authblk}
\usepackage{setspace}
\usepackage{comment}
\usepackage{setspace}
\usepackage{amsmath,amssymb,amsfonts,amsthm,mathrsfs}
\usepackage{graphicx}
\usepackage{enumerate} 
\usepackage[pdftex]{hyperref}

\def\be{\begin{equation}}
\def\ee{\end{equation}}
\def\ba{\begin{eqnarray}}
\def\ea{\end{eqnarray}}
\def\bi{\begin{itemize}}
\def\ei{\end{itemize}}

\def\bra{\langle}
\def\ket{\rangle}

\def\out{{\rm out}}
\def\in{{\rm in}}
\def\O{\Omega}
\def\xh{\hat{x}}
\def\ph{\hat{p}}
\def\I{\mathcal{I}}
\def\e{\varepsilon}
\def\qh{\hat{q}}
\def\yh{\hat{y}}
\def\wb{\bar{w}}
\def\zb{\bar{z}}

\def\t{\tau}
\def\H{\mathcal{H}}
\def\Qs{Q^{\rm soft}}
\def\Qh{Q^{\rm hard}}
\def\Qhh{\hat{Q}^{ \rm hard}}
\def\G{\Gamma}
\def\Gm{\Gamma_m}
\def\Gg{\Gamma_{\text{grav}}}
\def\scri{\mathcal{I}}
\def\fh{f_\H}
\def\xio{\mathring{\xi}}
\def\O{\Omega}
\def\Og{\Omega_{\text{grav}}}
\def\Om{\Omega_{m}}
\def\tf{\text{TF}}
\def\s{\, {\rm s}}
\def\ft{\tilde{f}}

\def\st{  \text{ST}}
\def\Vh{V_\H}
\def\L{\mathcal{L}}

\def\Vt{\tilde{V}}
\def\S{\mathcal{S}}
\def\J{\mathbf{J}}
\def\grav{\text{grav}}

\title{Asymptotic symmetries of gravity and soft theorems for massive particles}
\author[a]{Miguel Campiglia}
\author[b]{Alok Laddha}
\affil[a]{Universidad de la Rep\'ublica, Montevideo, Uruguay}
\affil[b]{Chennai Mathematical Institute, Chennai, India}

\begin{document}
\maketitle
\thispagestyle{empty}

\let\oldthefootnote\thefootnote
\renewcommand{\thefootnote}{\fnsymbol{footnote}}
\footnotetext{Email: campi@fisica.edu.uy, aladdha@cmi.ac.in}
\let\thefootnote\oldthefootnote

\begin{abstract}
The existing equivalence between (generalized) BMS Ward identities with leading and subleading soft graviton theorems is extended to the case where the scattering particles are massive scalars.  By extending the action of generalized BMS group off null infinity at late times, we show that there is a natural action of such group not only on the radiative data at null infinity but also on the scattering data of the massive scalar field.  This leads to a formulation of Ward identities associated to the generalized BMS group when the scattering states are massive scalars or massless gravitons and we show that these Ward identities are equivalent to the leading and subleading soft graviton theorems.

\end{abstract}

\section{Introduction}
Since the seminal work of Strominger \cite{strom1,stromBMS1}, there has been a resurgence of interest in analyzing the role played by BMS group \cite{aareview} and its various extensions \cite{barnich1} as symmetry group of (perturbative) Quantum Gravity S-matrix \cite{stromST,virasoro}. However work done so far is restricted to pure gravity or pure gravity coupled to massless matter. In this paper, we extend the relationships between a group of asymptotic symmetries ${ \cal G}$ defined in \cite{cl1,cl2} (referred to as generalized BMS group) with leading and sub-leading soft graviton  theorems \cite{weinberg,cs}.\\
Our work is based on the ideas presented in \cite{cl3}, where the Weinberg soft photon theorem for scalar QED with massive charged particles was shown to be equivalent to Ward identities associated to large $U(1)$ gauge transformations which acted on the asymptotic phase space of massive scalar field defined at time-like infinity. In this paper we show that there exists an action of ${\cal G}$ on the asymptotic phase space of massive scalars such that charges associated to ${\cal G}$ are sum of charges associated to time-like infinity and null infinity.\footnote{For the supertanslation subgroup $\st \subset \cal G$,  the action on the massive scalar seems to agree with the one proposed by Longhi and Materassi \cite{longhi} (see also \cite{gomis}).} We then show that the statement of Ward identities corresponding to such charges is equivalent to leading as well as sub-leading soft graviton theorems.\\
A key player in obtaining this equivalence are certain `boundary to bulk' Green's functions that map generators of ${\cal G}$ (which are vector fields at null infinity) to the asymptotic (at time-like infinity) bulk vector fields. We show that these Green's functions are intricately tied to the soft factors which arise in the soft theorems and this fact plays the central role in obtaining the equivalence between soft theorems and Ward identities.  Further details of these Green's functions appear in the companion note \cite{appendix}. \\
The outline of this paper is as follows. After giving a brief conceptual sketch of the key ideas which underlie the proof (establishing equivalence between Ward identities and soft theorems), in section \ref{section2} we  establish the asymptotic phase space $\G$ of the scalar field-gravity system. This phase space is a direct product of radiative phase space of gravity defined at null infinity and the asymptotic phase space of the massive scalar field defined at  time-like infinity. In section \ref{section3} we show that there is a group of large diffeomorphisms which are non-trivial at time-like infinity and are (i) obtained from the generalized BMS vector fields localized at null-infinity with help of certain Green's function with well-defined boundary conditions and  (ii) preserve the de Donder gauge-fixing condition of perturbative gravity. This gives us a unified picture of the asymptotic symmetry group of gravity associated to null and time-like infinity. 
In section \ref{section4} it is shown that the action of this symmetry group on the total phase space $\G$ is symplectic and we derive the charges associated to the symmetry generators.  In sections \ref{section5} and \ref{section6} we establish the equivalence between soft theorems and Ward identities associated to such charges. We end with some remarks and conclusions.

\subsection{Basic sketch of the proof underlying the equivalence}

We now illustrate the sequence of steps which leads to the equivalence in the context of the supertranslation subgroup $\st \subset {\cal G}$ and Weinberg's soft graviton theorem.
It is important to recall that the underlying theory is perturbative quantum gravity coupled to a massive scalar field where we use de Donder gauge to describe the linearized metric.

Let us for concreteness focus on future asymptotics.  We use two different systems of coordinates: $(u,r,\hat{x})$  adapted to future null infinity $\I^+$ and  $(\tau,\rho,\hat{x})$ adapted to future  time-like infinity $\H^+$. Null infinity  is reached by taking $r \to \infty$ in the first coordinates and  time-like infinity  is reached by taking $\t \to \infty$ in the second coordinate system.

Given a supertranslation vector field $f\partial_{u}$ at $\I^+$, we can consider an associated spacetime `bulk' vector field $\xi_{f}$ that is a residual `gauge' symmetry in de Donder gauge, i.e. satisfies $\square\xi_{f}^{a}\ =\ 0$ and its boundary value at ${\cal I}^{+}$ is $f\partial_{u}$.  At time-like infinity such vector field has the asymptotic form $\xi^\t_f = \xio^\t_f \partial_\t +O(\t^{-1})$ and its leading component can  be determined via a Green's function $G(\rho,  \xh; \hat{y})$ that maps $f(\hat{y})$ to $\xio^\t_f(\rho,\hat{x})$ by:
\be
 \xio^\t_f(\rho,\hat{x}) = \int_{S^{2}} d^2 \yh \, G(\rho,\xh;\hat{y})f(\hat{y}).
 \ee

If we now consider a scattering amplitude involving $n$ massive particles then the corresponding Ward identity for supertranslation \cite{stromBMS1},
\begin{equation}
\begin{array}{lll}
\langle \out\vert [\Qs, \S] \vert \in \rangle\ =\ -\langle \out \vert [\Qh,\S]\vert\in \rangle
\end{array}
\end{equation}
takes the form (for details see the main text of the paper):
\begin{multline}
\lim_{E_s \to 0^+} \frac{E_s}{2 \pi} \int d^2 w  f(w,\wb)   D^2_{\wb} \bra {\rm out} |a_+(E_s,w,\wb) \S | {\rm in} \ket  =\\
 -  \sum_{i=1}^{n} m_i  \int d^2 \yh \, G(\vert \vec{p}_{i}  /m_i\vert ,\hat{p}_{i}; \yh)  \, f(\yh) \bra {\rm out} |  \S | {\rm in} \ket .
\end{multline}
It is important to notice here that in the   $\tau\rightarrow \infty$ limit the momentum of a free massive particle $\vec{p}$ determines the  point on $\H^+$ the particle reaches via $\rho=\vert \vec{p}/m\vert,\ \hat{x}=\hat{p}$.
 
This formula looks structurally similar to Weinberg's soft graviton theorem which is given by
\be
\lim_{E_s \to 0^+} E_s  \bra {\rm out} | a_{+} (E_s,w,\wb) \S | {\rm in} \ket  = \sum_{i=1}^n \frac{(\e^+ \cdot p_i)^2}{(q/E_s) \cdot p_i} \bra {\rm out} |  \S | {\rm in} \ket
\ee
We show  below that after expressing boundary coordinate $\hat{y}$  into sphere coordinates $(w,\wb)$, one has the following remarkable relation between Green's function associated to supertranslation generators and soft factor
\begin{equation}
- m_i \, G(\vert \vec{p}_{i}/m_i\vert,\hat{p}_{i}; w,\wb) = \frac{1}{2 \pi} D_{\wb}^{2}\frac{(\e^+ \cdot p_i)^2}{(q/E_s) \cdot p_i}. \label{keyrel}
\end{equation}
This relation is the key in establishing the equivalence between ST Ward identities and Weinberg's theorem.  Similar sequence of logic interspersed with   Green's function associated to sphere vector fields on the boundary and certain derivatives of sub-leading soft factor leads to the equivalence between Ward identities associated to $\textrm{Diff}(S^{2}) \subset { \cal G}$ and Cachazo-Strominger (CS) subleading soft theorem.

We conclude by commenting  on the relationship with the case of massless particles. The equivalence  in such case can be cast in the above language by expressing the supertranslation action on massless particles in  terms of a `Green's function' that is just an identity kernel. The analogue of relation (\ref{keyrel}) is then:
\be
- E_i \, \delta^{(2)}(z_i,w) =  \frac{1}{2 \pi} D_{\wb}^{2}\frac{(\e^+ \cdot p_i)^2}{(q/E_s) \cdot p_i} \quad  \text{(massless particles)},  \label{keyrel2}
\ee
where $(z_i,\zb_i)$ are sphere coordinates for $\ph_i$.    The $\rho \to \infty$  boundary behavior obeyed by the Green's function $G(\rho,\xh;\yh)$ ensures that  Eq. (\ref{keyrel})  reduces to Eq. (\ref{keyrel2}) in the  $m \to 0, \;  \vec{p}= $ constant limit.


\section{Asymptotic phase space} \label{section2}
We consider perturbative gravity coupled to a massive scalar field in de Donder gauge. We assume a total phase space of the form 
$\Gamma\ =\ \Gg\times\Gamma_{m}$ with $\Gg$ and $\Gm$ the free-field asymptotic phase spaces of gravity and massive field respectively. 
\subsection{Gravity phase space $\Gg$}
The gravity phase space is described by the `radiative data' $C_{AB}(u,\xh)$ at null infinity  with symplectic structure\footnote{$C_{AB}(u,\xh)$ is assumed to satisfy $C_{AB}(u,\xh) = C^\pm_{AB}(\xh) + O(|u|^{-\epsilon})$ as $u \to \pm \infty$.}
\be
\Og(\delta,\delta') = \frac{1}{4}\int_{\scri} du \sqrt{\gamma} \left(\delta C^{AB} \delta' \dot{C}_{AB}  - \delta \leftrightarrow \delta' \right)  . \label{ssgrav}
\ee
In quantum theory this data is related to the asymptotic Fock functions as follows. Consider the Fourier transform of $C_{AB}$,
\be
C_{AB}(E,\xh) := \int_{-\infty}^\infty  C_{AB}(u,\xh) e^{i E u} du ,
\ee
and go to $(z,\zb)$ coordinates on the sphere. Then for $E>0$ the positive and negative helicity graviton anihilation functions of momentum $\vec{p}= E \xh$ are:
\be
a_+(E,\xh) = \frac{2 \pi i }{\sqrt{\gamma}} C_{zz}(E,\xh), \; \quad a_-(E,\xh) = \frac{2 \pi i }{\sqrt{\gamma}} C_{\zb\zb}(E,\xh), \label{aC} 
\ee
where $\sqrt{\gamma}=2/(1+z \zb)^2$. The symplectic structure (\ref{ssgrav}) implies the standard linearized gravity Poisson brackets:
\be
\{a_h(\vec{p}), a^*_{h'}(\vec{p}') \} = -i 2 E_{\vec{p}} \, \delta_{h h'} (2 \pi)^3   \delta^{(3)}(\vec{p}-\vec{p}').
\ee
\subsection{Massive scalar field phase space $\Gm$} \label{gammam}
The massive scalar field phase space is described in terms of data on a unit hyperboloid $\H$ describing time-like infinity. The coordinates adapted to such description are 
\be
\t := \sqrt{t^2 - r^2} , \quad  \rho  :=  \frac{r}{\sqrt{t^2 - r^2}} ,
\ee
in terms of which the Minkowski metric reads:
\be
ds^2 = - d \t^2 + \t^2  d \sigma^2, \label{metric}
\ee
with
\be
d \sigma^2  =  \frac{d \rho^2}{1+ \rho^2} + \rho^2 \gamma_{AB} d x^A d x^B   \; = : h_{\alpha \beta} d x^\alpha d x^\beta
\ee
the unit hyperboloid metric (with scalar curvature  -6).
At large $\tau$ the free massive field behaves as:
\be
\varphi(\t,\rho,\xh) = \frac{\sqrt{m}}{2 (2 \pi \tau)^{3/2}} \left(   b(\rho,\xh) e^{ -i \t m } + b^*(\rho,\xh) e^{ i \t m } \right)+O(\t^{-5/2}) , \label{asymphi}
\ee
with $b(\rho,\xh)$ representing free data.  To ensure well-definedness of upcoming expressions involving integrals in $\H$, we will assume the free data satisfies `finite energy' $\rho \to \infty$ fall-offs:
\be
b(\rho,\xh) = O(\rho^{-3/2-\epsilon}). \label{falloffb}
\ee
The symplectic structure is
\be
 \Om(\delta,\delta') = \frac{i m^2}{2 (2 \pi)^3} \int_{\H} d^3V \left( \delta b \, \delta' b^*- \delta \leftrightarrow \delta' \right). \label{ssphi}
\ee
In quantum theory $b(\rho,\xh)$ become the anihilator operator of a scalar particle with momentum $\vec{p}= \rho \xh$.   The symplectic structure (\ref{ssphi}) implies the standard  Poisson brackets:
\be
\{b(\vec{p}),b^*(\vec{p}') \}  = -i  (2 \pi)^3 (2 E_p) \delta^3(\vec{p}-\vec{p}').
\ee

\section{Extension of $\cal G$ to time-like infinity} \label{section3}
The residual gauge transformations in de Donder gauge are generated by vector fields satisfying the wave equation (with respect to the fixed reference Minkowski metric):
\be
\square \xi^a =0 . \label{boxxi}
\ee
We are interested  in  `large' gauge transformation that are non-trivial at infinity. At null infinity we would like to have generalized BMS vector fields. These vector fields are defined by the condition of being asymptotically divergence-free at null infinity \cite{cl1}: 
\be
\nabla_a \xi^a =O(r^{-1}). \label{divzero}
\ee
These vector fields are parametrized by sphere functions $f$ (supertranslations) and sphere vector fields $V^A$(generalized rotations) according to,
\be
\xi^a(r,u,\xh) = f \partial_u +V^A \partial_A + u \alpha \partial_u - r \alpha \partial_r + \ldots , \label{xibms}
\ee
where $2 \alpha$ is the 2-d divergence of $V^A$ and the dots indicate subleading terms in the $1/r$ expansion. In the usual treatment these terms  are determined by the Bondi gauge condition:
\be
g_{rr}=g_{rA}=0, \quad \det g_{AB}  = r^{4} \det \mathring{q}_{AB} ,
\ee
where $\mathring{q}_{AB}$ is the unit round sphere metric (in spherical coordinates, $\det \mathring{q}_{AB}= \sin^2 \theta$).
However in the present case one should instead use condition (\ref{boxxi}) to fix such subleading term (see \cite{avery} for such determination in the case of supertranslations). For the purposes of the present paper we will  not need the specific form of such subleading terms. 

We start with the following ansatz for the asymptotic expansion of the vector fields off time-like infinity 
\ba
\xi^\t(\t,\rho,\xh) & = & \xio^\t (\rho,\xh) +O(\t^{-1}) \label{xitau}  \\
 \xi^\alpha(\t,\rho,\xh) & = & \xio^\alpha(\rho,\xh)+ O(\t^{-1}) \label{xialpha}
\ea
where $\alpha,\beta,\dots$ denote indices on the hyperboloid.  As we will see shortly, this ansatz is consistent with Eqs. (\ref{boxxi}), (\ref{divzero}), (\ref{xibms}).  Note also that a vector field satisfying (\ref{xitau}), (\ref{xialpha}) has a well-defined action on the massive field free data $b(\rho,\xh)$ (obtained by evaluating the derivative $\xi^a \partial_a \varphi$ in the expansion (\ref{asymphi})) given by:
\be
\delta_\xi b = - i m \xio^\t b + \xio^\alpha \partial_\alpha b .\label{deltaxib}
\ee

The idea is  to use  (\ref{boxxi}), (\ref{divzero}), (\ref{xibms}) to determine $\xio^\t$ and $\xio^\alpha$
in terms of $f$ and $V^A$. In appendix \ref{xiapp} it is shown that conditions (\ref{boxxi}), (\ref{divzero}), (\ref{xibms}) imply:
\be
\Delta \xio^\t = 3\xio^\t   ,\quad  \quad \lim_{\rho \to \infty} \rho^{-1} \xio^\t (\rho,\xh) = f(\xh)  , \label{eqxiot}
\ee
\be
\Delta \xio^\alpha = 2 \xio^\alpha, \quad D_\alpha \xio^\alpha =0, \quad \quad \lim_{\rho \to \infty} \xio^A(\rho,\xh) = V^A(\xh) , \label{eqxioalpha}
\ee
where $\Delta$ and $D_\alpha$ are the Laplacian and covariant derivative on $\H$. If $G(\rho,\xh; \qh)$ and $G^\alpha_A(\rho,\xh; \qh)$ are  Green's functions for equations (\ref{eqxiot}) and (\ref{eqxioalpha}) respectively,  the solutions can be written as:
\be
\xio^\t(\rho,\xh) = \int_{S^2} d^2 \qh \, G(\rho,\xh; \qh) f(\qh) = : \fh(\rho,\xh) ,\label{deffh}
\ee
\be
\xio^\alpha(\rho,\xh) = \int_{S^2} d^2 \qh \, G^\alpha_A(\rho,\xh; \qh) V^A(\qh) =: \Vh^\alpha(\rho,\xh) . \label{defvh}
\ee
The explicit form of $G(\rho,\xh; \qh)$ and $G^\alpha_A(\rho,\xh; \qh)$   is given below in Eqns. (\ref{greenst2}) and (\ref{greenV2}) respectively.  Further details of these Green's functions are given in \cite{appendix}. 

The above extension of generalized BMS vector fields to time-like infinity provides  an action of generalized BMS on the phase space of massive particles and hence on the total phase space $\G= \Gg \times \Gm$. We now describe the associated charges.  

\section{Action of ${\cal G}$ on $\Gamma$}\label{section4}
In this section, we derive the complete expression for charges associated to all the generators of ${\cal G}$. As $\Gamma$ is the Cartesian product of radiative phase space of gravity defined at ${\cal I}$ and asymptotic phase space of scalar field defined at $\H$, charges associated to any (generalized) BMS vector field are a sum of gravitational charges defined at $\I$ \cite{stromBMS1,cl1}, and scalar charges defined at $\H$. Exactly as in the case of pure gravity \cite{stromST,cl2}, charges associated to supertranslation and sphere vector fields split into contributions which can be classified as hard and soft charges. The soft charge is linear in an infinite wavelength mode of $C_{AB}$. As the gravitational charges associated to the radiative phase space $\Gg$ were derived in earlier papers, we do not reproduce the derivation here and only derive the charges associated to scalar field which are functions on $\Gamma_{m}$.

\subsection{Supertranslation charges}
From the previous consideration, we have an action of supertranslations on the total phase space $\G=\Gg \times \Gm$:
\ba
\delta_f C_{AB} & = & f \partial_u C_{AB} - 2 (D_A D_B f)^{\tf}  \label{deltafC} \\
\delta_f b & =&  - i m \fh b,  \label{deltafb2}
\ea
with $\fh$ defined in (\ref{deffh}). The generator $Q_f$ satisfying $\delta Q_f = \O(\delta,\delta_f)$ can be written as
\be
Q_f =  \Qh_f + \Qs_f. \label{Qf}
\ee
The hard part is a sum of gravitational and matter contributions
\be
\Qh_f=(\Qh_f)_\grav + (\Qh_f)_{\rm matter}.
\ee
The gravitational contribution was derived in \cite{AS} and is given by
\ba
(\Qh_f)_\grav & = &  \frac{1}{2}\Og(f \partial_u C, C) \\
&= & \frac{1}{4}\int_{\scri} du \sqrt{\gamma} f \partial_u C^{AB} \partial_u C_{AB}.
\ea
The matter contribution is given by:
\ba
 (\Qh_f)_{ \rm matter}  & = & \frac{1}{2} \Om( \delta_f b,b) \\
 & = & \frac{m^{3}}{2(2\pi)^{3}}\int_{{\cal H}} d^3 V f_{{\cal H}} b^* b. \label{Qfb}
\ea
One can verify that (\ref{Qfb})  coincides with the expected expression in terms of the energy-momentum tensor: 
\be
(\Qh_f)_{ \rm matter} = - \lim_{\t \to \infty} \int_{\H_\t} d S_a \sqrt{g} \, T^a_b \xi^b_f ,\label{Qfb2}
\ee
where $\H_\t$ is the $\t=$constant hypersurface and  $\xi^a_f \partial_a = \fh \partial_\t+ O(\t^{-1})$. 
For the translation subgroup of supertranslations, Eq.  (\ref{Qfb2}) gives the  total linear momentum of the field (in the asymptotic future).

We now describe the soft part in (\ref{Qf}). This was derived in \cite{AS}, and can be written as:
\be
\Qs_f =  \Og(- 2 D^2 f, C) = -\frac{1}{2} \int_{S^2} d^2 V f D^A D^B [C_{AB}]. \label{Qsf}
\ee
Where the square bracket denotes difference of boundary values at $u = \pm \infty$:
\be 
[C_{AB}](\xh) := C_{AB}(\infty,\xh)- C_{AB}(-\infty,\xh).
\ee
In order to establish the equivalence between ST Ward identities and Weinberg's soft theorem we need to impose the additional condition:\footnote{Conversely, Weinberg's soft theorem can be seen to imply condition (\ref{magzero}).}
\be
D^z D^z [C_{zz}]= D^{\zb} D^{\zb} [C_{\zb\zb}]. \label{magzero}
\ee
This condition is satisfied by the so-called Christodolou-Klainermann space-times considered by Strominger in \cite{strom1}.\\
We conclude the section by writing the soft charge in terms of the mode functions. First, we express $[C_{AB}]$ as a zero energy limit of the Fourier transform $C_{AB}(E,\xh)$:
\be
[C_{AB}(\xh)] = -i \lim_{E \to 0^+}  E \, C_{AB}(E,\xh) . \label{bdyC}
\ee
Using (\ref{aC}) and (\ref{magzero}) the soft charge (\ref{Qsf}) can be written as:
\be
\Qs_f = \frac{1}{2 \pi} \lim_{E \to 0^+} E \int d^2 z f  D^2_{\zb}  a_+(E,z,\zb) \label{Qsfa}
\ee
(or alternative expression  in terms of  $D^2_z a_-$). A technical but important point is that when performing covariant derivatives as in (\ref{Qsfa}) one needs to remember the two-dimensional tensor structure (including density weight) of the quantity being derived. For the case of $a_+$ this tensor structure can be read off from Eq. (\ref{aC}). Doing so one finds that the differential operator in (\ref{Qsfa}) acts as:
\be
D^2_{\zb}  a_+(E,z,\zb)  =  \partial_{\zb}\big( \gamma^{z \zb} \partial_{\zb} (\gamma_{z \zb} a_+(E,z,\zb))\big). \label{D2itopz}
\ee

\subsection{Sphere vector field charges}
In this section we derive the charges associated to the generators of $\textrm{Diff}(S^{2})\ \subset\ {\cal G}$. The corresponding generators of ${\cal G}$ are vector fields at null infinity which are in turn parametrized by vector fields $V^{A}\partial_{A}$ on the conformal $S^{2}$.
\begin{equation}
\xi^{a}(u,\hat{x})\partial_{a} =\ V^{A}\partial_{A}\ +\ u\alpha\partial_{u},
\end{equation}
$\alpha\ =\ \frac{1}{2}(D_A V^A)$.\\
For these vector fields their action on $\Gamma$ is given by,
\ba
\delta_V C_{AB} & = & \delta_V^{\text{hard}}C_{AB} + \delta_V^{\text{soft}}C_{AB}   \label{deltaVC} \\
\delta_V b & =&  \L_{\Vh} b,  \label{deltaVb}
\ea
where
\ba
\delta_V^{\text{hard}}C_{AB} & := & \L_V C_{AB} - \alpha C_{AB} + \alpha u \partial_u C_{AB}   \label{dVh}\\
 \delta_V^{\text{soft}}C_{AB}  & :=  &- 2 u (D_A D_B \alpha)^{\tf}
\ea
and $\Vh$ defined in (\ref{defvh}). There is a subtlety in the present case in that strictly speaking the action (\ref{deltaVC}) is between radiative phase spaces associated with different 2-dimensional metrics $q_{AB}$. As shown in \cite{cl2}, one can nevertheless compute the associated charge by embedding $\Gg$ into a larger space that allows for variation of $q_{AB}$. Doing so one finds the charge is  a sum of `hard'  and `soft'  pieces. The `hard' piece turns out to  coincide with the naive expression,
\ba
(\Qh_{V})_{\text{grav}} & = & \frac{1}{2}\Og(\delta_V^{\text{hard}}C,C)\\
&=&\frac{1}{4} \int_\I du  \sqrt{\gamma} \,\partial_u C^{AB}(\L_V C_{AB} - \alpha C_{AB} + \alpha u \partial_u C_{AB}). \label{QhVg}
\ea
On the other hand, the soft charge receives contributions from the variation of the 2-metric $\delta_V q_{AB}$ resulting in,\footnote{This charge is only defined on the subspace of $\Gg$ satisfying the stronger fall-offs $C(u,\xh)=O(|u|^{-1-\epsilon})$ at $u \to \pm \infty$. The charge  \ref{defQsV} was first given in \cite{virasoro} for the case where $V^A$ is local conformally Killing. In such case $\delta_V q_{AB}=0$ and the charge can be derived within $\Gg$.}
\be
\Qs_{V} =  \frac{1}{2} \int_{\I} du  \sqrt{\gamma}  \, (C^{zz}  D^3_z V^z +C^{\zb \zb}  D^3_{\zb} V^{\zb})  ,\label{defQsV}
\ee
which differs from the naive expression $ \Og(\delta_V^{\text{soft}}C,C)$.

For the matter contribution, we note that the fact that $\Vh^\alpha$ is divergence-free as shown in Eq.(\ref{eqxioalpha}) implies the action (\ref{deltaVb}) is symplectic on $\Gm$.\footnote{Here we are crucially using the fact that the  field is a scalar. For fields with nonzero spin the symplectic form will depend on the full hyperboloid metric $h_{\alpha \beta}$ (not just its volume element $\sqrt{h}$) and further subtleties will  arise.} The associated charge is then given by
\ba \label{QhVm}
(\Qh_V)_m & := & \frac{1}{2} \Om( \delta_V b,b)\\
& = &  \frac{im^{2}}{2(2\pi)^{3}}\int_{{\cal H}}d^{3}V \, b^{*} \L_{\Vh} b. \label{QVb}
\ea
One can verify (\ref{QVb})  coincides with the expected expression from the energy-momentum tensor perspective. In the notation of Eq. (\ref{Qfb2}): 
\be
(\Qh_V)_m = - \lim_{\t \to \infty} \int_{\H_\t} dS_a T^{a}_b \xi^b_V ,
\ee
where $\xi^b_V \partial_b= \Vh^\alpha \partial_\alpha + O(\t^{-1})$.  For the vector fields  associated  to rotations and boosts,  Eq.  (\ref{Qfb2}) gives the total angular momentum of the field (in the asymptotic future).

The total sphere vector field charge is then given by
\be
Q_V = \Qh_V + \Qs_V
\ee
with the hard piece $\Qh_V \equiv (\Qh_{V})_{\text{grav}} + (\Qh_V)_m$ given by Eqns. (\ref{QhVg}), (\ref{QVb}) and the soft piece  given by Eq. (\ref{defQsV}).
We conclude the section by writing the soft charge in terms of the mode functions.  Using (\ref{aC}) and the prescription given in \cite{virasoro} that projects out the Weinberg pole, $\int du \, C(u,\xh) = \lim_{E \to 0} (1+ E \partial_E) C(E,\xh)$, the soft charge can be written as:
\be
\Qs_V = - \frac{1}{4 \pi i} \lim_{E \to 0^+} (1+ E \partial_E) \int d^2 z \big(V^{\zb} D^3_{\zb} a_+(E,z,\zb)+V^{z} D^3_z a_-(E,z,\zb) \big). \label{QsV}
\ee
Taking into account the tensorial structure of the mode functions, the explicit action of the differential operators in (\ref{QsV}) is found to be given by:
\be
D^3_{\zb} a_+ = \partial^3_{\zb} a_+, \quad  D^3_{z} a_- = \partial^3_{z} a_- .
\ee

\section{Supertranslation Ward Identity $\equiv$  Weinberg soft theorem} \label{section5}
Having derived the action of generalized BMS group on the gravity-massive scalar field asymptotic phase space, we now turn to the quantum theory and analyze the constraints such a symmetry imposes on the perturbative S-matrix of the theory. That is, we ask that if indeed the generalized BMS group was a symmetry group of the perturbative S-matrix,  what would be its implications.\\
As in the case of gravity coupled to massless particles \cite{stromST}, in this section we show that the infinity of  Ward identities associated to supertranslation subgroup are equivalent to the Weinberg's soft graviton theorem. 
We first review the equivalence of supertranslation Ward identities with the Weinberg soft theorem in the case of pure gravity and then generalize it to our case.
\subsection{Review for the case of external massless particles}
To setup notation for the later section we review the equivalence for the case of external massless particles. 
For given `in' and `out' states composed of massless particles of momenta $\{ \vec{p}_i \}$, the Ward identity associated to a supertranslation $f$,
\be
\bra \out | Q_f \S - \S Q_f | \in \ket=0 \label{wardidf}
\ee
 can be written as
\begin{multline}
 \lim_{E_s \to 0^+} \frac{E_s}{2 \pi} \int d^2 w  f(w,\wb)   D^2_{\wb}  \bra {\rm out} |a_+(E_s,w,\wb) \S | {\rm in} \ket  = \\ - \sum_i  E_i  f(\ph_i)\bra {\rm out} |  \S | {\rm in} \ket .  \label{wizrm}
\end{multline}
In (\ref{wizrm}) the sum is over all external particles, with $E_i = \pm | \vec{p}_i |$ for outgoing/incoming particles and $\ph_i=\vec{p}_i/|\vec{p}_i|$. 

On the other hand, Weinberg's soft graviton theorem can be written as:\footnote{Here and below, the soft theorems are written with sign convention such that all particles are outgoing.}
\be
\lim_{E_s \to 0^+} E_s  \bra {\rm out} | a_{+} (E_s,w,\wb) S | {\rm in} \ket  = \sum_{i} \frac{(\e^+(w,\wb) \cdot p_i)^2}{(q/E_s) \cdot p_i} \bra {\rm out} |  S | {\rm in} \ket , \label{wstzrm}
\ee
where  $q/E_s \equiv (1,\qh)$ with $\qh$ parametrized by $(w,\wb)$ and $\e^+(w,\wb)$ the polarization vector \cite{stromST}:
\be
\e^{+ \mu}(w,\wb)=1/\sqrt{2}(\wb,1,-i,-\wb).
\ee
If we parametrize  an external momentum $\vec{p}$ by $(E,z,\zb)$, the soft factor in (\ref{wstzrm}) take the form:
\be
\frac{(\e^+(w,\wb) \cdot p)^2}{(q/E_s) \cdot p} = - E \s(z,\zb; w,\wb),
\ee
with
\be\label{august16-4}
\s(z,\zb; w,\wb):= \frac{1+ w \wb}{1+ z \zb} \frac{\wb-\zb}{w-z} .
\ee
To go from (\ref{wstzrm})  (soft theorem) to (\ref{wizrm}) (Ward identity for supertranslation $f$) one performs the operation operation $(2 \pi)^{-1} \int d^2 w   f(w,\wb)  D^2_{\wb} $ on both sides of (\ref{wstzrm}). The LHS becomes the left term in (\ref{wizrm}). That the RHS also coincides follows from the identity (see appendix \ref{D2softapp}):
\be
 D^2_{\wb} \s(z,\zb; w,\wb) =  2 \pi  \delta^{(2)}(w-z) .\label{D2soft}
\ee
To go from (\ref{wizrm}) to (\ref{wstzrm}) we look at the Ward identity for the particular function
\be
f(z,\zb) =  \s(z,\zb; w,\wb).
\ee
The RHS becomes the right term in (\ref{wstzrm}). That the LHS also coincides follows from an integration by parts and the identity (see appendix \ref{D2softapp}):
\be
D^2_{\zb}   \s(z,\zb; w,\wb) =  2 \pi  \delta^{(2)}(w-z) .  \label{D2soft2}
\ee

\subsection{External massive particles}
By using normal ordered prescription to define the quantum charge $\Qhh_f$ one has,
\be
[\hat{b}(\vec{p}), \Qhh_f] = m \fh(\vec{p}/m) \hat{b}(\vec{p}) \label{bQfcomm}
\ee
(see Eq. (\ref{deltafb2})).  Using (\ref{bQfcomm}), the proposed Ward identity (\ref{wardidf}) for external massive scalars takes the form:
\begin{multline}
 \lim_{E_s \to 0^+} \frac{E_s}{2 \pi} \int d^2 w  f(w,\wb)   D^2_{\wb}  \bra {\rm out} |a_+(E_s,w,\wb) \S | {\rm in} \ket  =  \\-  \sum_i m_i  \fh(\vec{p}_i/m)  \bra {\rm out} |  \S | {\rm in} \ket ,  \label{wim}
\end{multline}
where $m_i = \pm m$ for outgoing/incoming particles. On the other hand, Weinberg's soft graviton theorem takes the same form as in Eq. (\ref{wstzrm}):
\be
\lim_{E_s \to 0^+} E_s  \bra {\rm out} | a_{+} (E_s,w,\wb) \S | {\rm in} \ket  = \sum_{i} \frac{(\e^+(w,\wb) \cdot p_i)^2}{(q/E_s) \cdot p_i} \bra {\rm out} |  \S | {\rm in} \ket . \label{wstm}
\ee
We now repeat the steps that led to the equivalence in the massless case. If we perfom the 
operation $(2 \pi)^{-1} \int d^2 w   f(w,\wb)  D^2_{\wb} $ on both sides of (\ref{wstm}) we obtain:
\be
 \lim_{E_s \to 0^+} \frac{E_s}{2 \pi} \int d^2 w  f(w,\wb)   D^2_{\wb}  \bra {\rm out} |a_+(E_s,w,\wb) \S | {\rm in} \ket  = -  \sum_i m_i  \ft(\vec{p}_i/m)  \bra {\rm out} |  \S | {\rm in} \ket ,  \label{wimp}
\ee
with
\be
\ft(\vec{p}/m) = \int d^2 w \, G(\vec{p}/m; w,\wb)  \, f(w,\wb), \label{fp}
\ee
\be
 G(\vec{p}/m; w,\wb) :=- \frac{1}{2\pi} D^2_{\wb} \frac{\big(\e^+(w,\wb) \cdot (p/m) \big)^2}{(q/E_s) \cdot (p/m)} . \label{greenst}
\ee
Parametrizing the 3-momentum particle as  $\vec{p}= m \rho \xh$ one can verify that (\ref{greenst}) takes the following simple form:
\be
G(\vec{p}/m ; \qh) =- \frac{1}{4 \pi} \frac{\sqrt{\gamma(\qh)}}{\left((q/E_s) \cdot (p/m)\right)^{3}}.\label{greenst2}
\ee
Now, by direct computation it can be verified (\ref{greenst2}) satisfies (see also \cite{appendix}):
\be
(\Delta_{(\rho,\xh)} - 3) G(\rho,\xh ; \qh)=0 .
\ee
Furthermore, using the fact that the $\rho \to \infty, m= $ constant limit can be written as a  $m \to 0$, $\vec{p} = $constant limit, together with Eq. (\ref{D2soft}) one finds:
\be
\lim_{\rho \to \infty} \rho^{-1} G(\rho,\xh ; \qh) = \delta^{(2)}(\xh,\qh).
\ee 
It then follows that $G$ is the Green's function for Eq. (\ref{eqxiot}), and so $\ft=\fh$.  Thus the identity (\ref{wimp}) coincides with the Ward identity (\ref{wim}). 

To go from (\ref{wim}) to (\ref{wstm}) we repeat the steps as in the massless case: Consider the Ward identity for the particular function
\be
f(w,\wb) =  \s(w,\wb; z_s, \zb_s).
\ee
where $\s(w,\wb; z_s, \zb_s)$ is defined in Eq.(\ref{august16-4}). By the same argument as in the pure gravity case, the LHS then becomes the left term  in  (\ref{wstm}). The multiplicative term on the RHS is given by
\ba
-m \fh(\vec{p}/m) &= & - m \int d^2 w \, G(\vec{p}/m; w,\wb)  \,  \s(w,\wb; z_s, \zb_s)  , \label{stibp1} \\
&=&  m \frac{\big(\e^+(w,\wb) \cdot (p/m) \big)^2}{(q/E_s) \cdot (p/m)}, \label{stibp2}
\ea
where we used the definition (\ref{greenst}) of $G$, integrated by parts and used Eq. (\ref{D2soft2}). Thus one recovers the right term in (\ref{wstm}).

\section{Sphere vector field Ward identity $\equiv$ CS soft theorem} \label{section6}
We  now turn our attention to the relationship between Ward identities arising from the $\textrm{Diff}(S^{2})$ vector fields and  the generalization of Cachazo Strominger soft theorem to the case where the external particles are massive scalars. Such a theorem is a factorization formula which relates the  scattering amplitude involving scattering states that include massive scalars as well as one sub-leading soft graviton  in terms of certain soft factors and amplitudes involving only the external massive scalar particles. This formula has only been derived in the limit that masses of scattering particles is zero.  However following \cite{broedel} we will try to give a persuasive argument to the effect that such  sub-leading theorem continues to hold in the present case as well.  

After presenting such argument we will review the Ward identity--soft theorem equivalence for massless particles and finally establish the equivalence for massive particles.
\subsection{Subleading soft theorem for external massive scalars}
The set up is as follows. We consider scattering amplitudes containing external scalar particles and one graviton and are interested in the limit of such  scattering amplitudes given by $\lim_{E_{s}\rightarrow\ 0}\left[{\cal M}_{n+1}(p_{1},\dots,p_{n}, q_{s})\ -\  \frac{1}{E_{s}}S^{(0)}{\cal M}_{n}\right]$. 
Here $S^{(0)}$ is the Weinberg soft factor whose structural form $\sum_{i=1}^{n}\epsilon_{\mu\nu}\frac{p_{i}^{\mu}p_{i}^{\nu}}{p_{i}\cdot q}$ remains the same irrespective of whether $p_{i}$'s are time-like or null. Subleading soft theorems like CS theorem are statements which express the above quantity in terms of ${\cal M}_{n}$ in the sense that 
\be
\lim_{E_{s}\rightarrow\ 0}\left[{\cal M}_{n+1}(p_{1},\dots,p_{n}, q_{s})\ -\  \frac{1}{E_{s}}S^{(0)}{\cal M}_{n}\right]\ =\  S^{(1)}{\cal M}_{n}.
\ee
In the case that the scattering particles are massless, in a beautiful paper Broedel et. al. \cite{broedel} prove that by postulating the form of $S^{(1)}$ as a differential operators on the momentum space of scattering particles,\footnote{Their arguments in fact are more general and apply to scattering particles with spin as well (in which case the postulated form of $S^{(1)}$ involves differential operators in momentum space as well as differential operators on the space of polarization tensors).} 
 one could constraint their form severely by demanding that scattering amplitude to be Poincare invariant, gauge invariant and that these differential operators associated to $S^{(1)}$ split as a sum over external particles as $S^{(1)}\ =
\ \sum_{i}S^{(1)}_{i}(\epsilon(q),q,p_{i},\partial_{p_{i}})$. The last condition is what is referred to as the locality constraint and is a reasonable condition to require when we are working with tree level amplitudes.  As noticed in \cite{broedel} there is yet another non-trivial constraint the soft factors have to satisfy which is related to the fact that scattering amplitudes are distributions in the momentum space as there is always an overall momentum conserving $\delta$-distribution multiplicative factor in their definition. If we refer to the scattering amplitudes without the momentum conserving $\delta$-fn. factor as reduced scattering amplitude, then by assuming that the soft factors are the same for the reduced as well as un-reduced scattering amplitudes\footnote{This assumption was not needed for Broedel et. al. and this fact is actually a consequence of all the constraints we have listed above. However this fact is not relevant for the issues being addressed here and hence we ignore it for simplicity.}
it was shown that $S^{(1)}$ has to satisfy,
\begin{equation}\label{distribution}
[S^{(1)},\delta^{(4)}(P)]\ =\ S^{(0)}\left(q\cdot\partial_{P}\delta^{(4)}(P)\right).
\end{equation}
The most general form for $S^{(1)}\ =\ \sum_{i}S^{(1)}_{i}(\epsilon(q),q,p_{i},\partial_{p_{i}})$ can be postulated based on the following conditions: \\
\noindent{\bf (i)} Poincare invariance implies that $S^{(1)}$ has to be linear in polarization tensor tensor,\\
\noindent{\bf (ii)}  Dimensional analysis shows that $S^{(1)}$ must have mass dimension zero and\\
\noindent{\bf (iii)} As it is sub-leading  it must be invariant under $q\rightarrow\ \lambda\ q$ for scaling parameter $\lambda$.\\

 Condition {\bf (i)} in conjunction with locality requirement implies that 
\begin{equation}\label{functionalform}
\begin{array}{lll}
S^{(1)}\ =\ \sum_{i}\epsilon_{\mu\nu}\omega^{\mu\nu\rho\sigma}p_{i}^{\rho}\frac{\partial}{\partial p_{i\sigma}} +\  O((\frac{\partial}{\partial p})^{2}) \  +\ S^{(1)}_{\textrm{function}}
\end{array}
\end{equation}
where $S^{(1)}_{\textrm{function}}$ is a multiplicative factor which doesnot involve differential operators on momentum space.

If the scattering particles are massless, then the most general form for $\omega^{\mu\nu\rho\sigma}$ which is consistent with conditions ({\bf ii}), ({\bf iii}) above is \cite{broedel}
\begin{equation}\label{masslessomega}
\begin{array}{lll}
\omega^{\mu\nu\rho\sigma}\ =\ \sum_{i}\left[c_{1}^{i}\frac{p^{\mu}_{i}p^{\nu}_{i}q^{\rho}q^{\sigma}}{(q\cdot p^{i})^{2}}\ +\ c_{2}^{i}\frac{\eta^{\rho(\mu}p^{\nu)}_{i}q^{\sigma}}{(q\cdot p_{i})}\ +\ c_{3}^{i}\frac{\eta^{\sigma(\mu}p_{i}^{\nu)}q^{\sigma}}{q\cdot p_{i})}\ +\ c_{4}^{i}\eta^{\rho(\mu}\eta^{\nu)\sigma}\right] .
\end{array}
\end{equation}

As shown in \cite{broedel}, gauge invariance and distributional constraints imply that in the massless case (that is when $p_{i}\cdot p_{i}\ =\ 0$), 
$S^{(1)}_{function}$ is zero, the higher derivative operators are absent as distributional constraint implies that we require each of them to annihilate the delta-function and the remaining first order differential term precisely reduces to the Cachazo-Strominger soft factor. We will now show that this result remains uneffected even when we drop the $p_{i}\cdot p_{i}\ =\ 0$ condition.

As in massless case we postulate that the factorized sub-leading soft factor $S^{(1)}$ takes the form
\begin{equation}
\begin{array}{lll}
S^{(1)}\ =\ \sum_{i}S^{(1)}_{i}(\epsilon_{\mu\nu},q,p_{i},\partial_{i})\ +\ S^{(1)}_{\textrm{function}}
\end{array}
\end{equation}
where $S^{(1)}_{\textrm{function}}$ is multiplicative and does not involve differential operators. The first time involves linear derivative operators as well as operators that will be $O(\partial_{i}^{2})$. However exactly as in the massless case, the distributional constraint will imply that these higher order operators have to annihilate $\delta^{4}(\sum_{i}p_{i})$ and hence will be zero. Thus just as in the massless case, the functional form of $S^{(1)}_{i}(\epsilon_{\mu\nu},q,p_{i},\partial_{i})$ is given by the same expression that appear in Eq.(\ref{functionalform}).\\
 As the  vectors $p_{i}$ are not null and if we assume that all the  scattered particles have same mass $m$ then the most general form of $\omega^{\mu\nu\rho\sigma}$ consistent with conditions {\bf (ii)}, {\bf (iii)} stated above, and such that on contraction with $\epsilon_{\mu\nu}$ and $p_{i}^{\rho}\frac{\partial}{\partial p_{i}^{\sigma}}$ no two terms get repeated is given by
\begin{equation}\label{massiveomega}
\begin{array}{lll}
\omega^{\mu\nu\rho\sigma}\ =\\
\vspace*{0.1in}
\sum_{i}\left[\dots\ +\ c_{5}^{i}m^{2}\frac{\eta^{\mu\nu}q^{\rho}q^{\sigma}}{(p_{i}\cdot q)^{2}}\ +\ c_{6}^{i}\frac{p_{i}^{\mu}p_{i}^{\nu}q^{\rho}p^{\sigma}}{(p_{i}\cdot q)m^{2}}\ +\ 2c_{7}^{i}\frac{p^{(\mu}\eta^{\nu)\sigma}p^{\rho}}{m^{2}}\right]
\end{array}
\end{equation}
where $\dots$ indicate the terms inside the parenthesis in eq.(\ref{masslessomega}).\\
We now impose conditions of gauge invariance $S^{(1)}$ and show that all the terms in Eq.(\ref{massiveomega}) which were not present in Eq.(\ref{masslessomega}) necessarily vanish.
We immediately see that owing to traceless-ness of $\epsilon_{\mu\nu}$ the term associated to $c_{5}^{i}$ is absent.\\
Gauge invariance of scattering amplitude implies that we need to impose
\begin{equation}
\begin{array}{lll}
S^{(1)}(\epsilon_{\mu\nu}\ =\ \lambda_{(\mu}q_{\nu)})\ =\ 0
\end{array}
\end{equation}
for any gauge parameter $\lambda_{\mu}$. This condition gives us 
\begin{equation}\label{gaugeinveqn}
\begin{array}{lll}
\sum_{i}\left[(2c_{1}^{i}\ +\ c_{2}^{i}\ +\ c_{3}^{i})q^{\rho}q^{\sigma}\frac{\lambda\cdot p_{i}}{q\cdot p_{i}}\ +\ (c_{3}^{i}\ + c_{4}^{i})q^{\rho}\lambda^{\sigma}\ +\ (c_{2}^{i}\ +\ c_{4}^{i})q^{\sigma}\lambda^{\rho}\right.\\
\vspace*{0.1in}
+\left. c_{6}^{i}\frac{(p_{i}\cdot\lambda)}{m^{2}}q^{\rho}p_{i}^{\sigma}\ +\ c_{7}^{i}\frac{(\lambda\cdot p_{i})p^{\rho}q^{\sigma}\ +\ (p_{i}\cdot q)p^{\rho}\lambda^{\sigma}}{m^{2}}\right]\\
\vspace*{0.1in}
\hspace*{2.7in}p_{i}^{\rho}\frac{\partial}{\partial p_{i}^{\sigma}}{\cal M}_{n}\ =\ 0
\end{array}
\end{equation}

We thus have following set of equations which includes the set that was obtained in \cite{broedel}
\begin{equation}
\begin{array}{lll}
2c_{1}^{i}\ +\ c_{2}^{i}\ +\ c_{3}^{i}\ =\ 0\\
c_{3}^{i} + c_{4}^{i}\ =\ -(c_{2}^{i}\ +\ c_{4}^{i})\ =\ c\\
c_{6}^{i}\ =\ 0\\
c_{7}^{i}\ =\ 0
\end{array}
\end{equation}
The second  equation allows that some of the terms in eq.(\ref{gaugeinveqn}) vanish due to conservation of total angular momentum.\\
Gauge-invariance also shows that $S^{(1)}_{\textrm{function}}$ part vanishes. As we require $S^{(1)}_{\textrm{function}}$ to satisfy conditions {\bf (i)},{\bf (ii)},{\bf (iii)} listed above, its most general form is 
\begin{equation}
S^{(1)}_{\textrm{function}}\ =\ \sum_{i}c_{8}^{i}\epsilon_{\mu\nu}\frac{p_{i}^{\mu}p_{i}^{\nu}}{m^{2}}
\end{equation}
Clearly gauge invariance would imply that 
\begin{equation}
\sum_{i}c_{8}^{i}(\lambda\cdot p_{i})(q\cdot p_{i})\ =\ 0
\end{equation}
which for arbitrary $\lambda, q^{\mu}$ imply that $c_{8}^{i}\ =\ 0$.\\
Thus, the extra possible terms with coefficients $c_5, c_6, c_7, c_{8}$ all vanish and the soft factor takes the same form as in the massless case. That is,  due to Poincare invariance, locality, gauge invariance as well as distributional constraint, assuming that there is a factorization in the sub-leading soft limit, the corresponding 
 sub-leading soft factor takes the same functional form irrespective of the masses of the hard particles.
\subsection{Review for the case of external gravitons}
We first sketch the equivalence between sphere vector field Ward identities and CS soft theorem for the case of external gravitons. We refer to \cite{cl1} for details. The Ward identity that follows from the condition 
\be
\bra \out | Q_V \S - \S Q_V | \in \ket=0 \label{wardidV}
\ee
for and `in' and `out' states composed of gravitons of momenta $\{ \vec{p}_i \}$ can be written as:
\begin{multline}
- \frac{1}{4 \pi} \lim_{E_s \to 0^+} (1+ E_s \partial_{E_s})\\
 \hspace*{0.4in}\int d^2 w (   V^{\wb} \partial^3_{\wb} \bra {\rm out}| a_+(E_s,w,\wb) \S  | {\rm in} \ket +  V^w \partial^3_w \bra {\rm out}|  a_-(E_s,w,\wb) \S | {\rm in} \ket ) = \\  \sum_i \J^i_{V} \bra {\rm out} |  \S | {\rm in} \ket.  \label{wivg}
\end{multline}
Here $\J^i_{V}$ is a differential operator associated to $V^A$ that acts on the momentum variables of the $i$-th  graviton. Its form is obtained from the action (\ref{dVh}) on the mode functions (\ref{aC}) (see  Eq. (57) of \cite{cl1} for the explicit expression).

On the other hand, Cachazo-Strominger subleading soft theorem can be written as:
\be
\lim_{E_s \to 0^+} (1+ E_s \partial_{E_s})   \bra {\rm out} | a^{\rm out}_{+} (E_s,w,\wb) \S | {\rm in} \ket  = \sum_{i}  \frac{\e^+(w,\wb) \cdot p_i}{q \cdot p_i} \, \e^+_\mu(w,\wb) q_\nu  J^{\mu \nu}_i \bra {\rm out} |  \S | {\rm in} \ket , \label{csg}
\ee
with $\e^+$ and $q$ as in Eq. (\ref{wstzrm})  and with $J^{\mu \nu}_i$ the total angular momentum operator on the $i$-th graviton.\footnote{We follow the conventions used in \cite{broedel} where $J_{\mu \nu}= p_\mu \partial_\nu - p_\mu \partial_\nu +$ helicity terms.  There are sign errors in Eqns. (63) and (68) of \cite{cl1} that cancelled each other. With the aforementioned conventions the correct identity is the one given by  (\ref{softJK}), (\ref{defKp}), and $\J_V= V^A \partial_A + \ldots$ (see Eq. (57) of \cite{cl1} for the complete expression of $\J_V$).}  In \cite{cl1} we noticed that the soft factor in (\ref{csg}) can be written as:
\be
\frac{\e^+(w,\wb) \cdot p_i}{q \cdot p_i} \,  \e^+_\mu(w,\wb) q_\nu J^{\mu \nu}_i = \J^i_{K^{+}_{(w,\wb)}} , \label{softJK}
\ee
where $\J^i_{K^{+}_{(w,\wb)}}$ is the differential operator associated to  the sphere vector field
\be
K^+_{(w,\wb)}  := \frac{(\zb-\wb)^2}{(z-w)}\partial_{\zb}  .\label{defKp}
\ee
 Similarly the factor associated to a negative helicity subleading  soft graviton can be written as the differential operator associated to the vector field $K^-_{(w,\wb)} $ that is the complex conjugated of (\ref{defKp}).

In order to go from (\ref{wivg}) to (\ref{csg}) we look at the Ward identity (\ref{wivg}) for the particular vector field
\be
V^A =  K^+_{(w,\wb)}.
\ee
By the identity (\ref{softJK}) the RHS  becomes the right term in (\ref{csg}). That the LHS also coincides follows from an integration by parts together with
\be
 \partial^3_{\zb} \frac{(\zb-\wb)^2}{(z-w)} =  4 \pi \delta^{(2)}(w-z). \label{d3K}
 \ee
Similarly one can obtain the negative helicity soft theorem by looking at the Ward identity associated to the vector field $K^-_{(w,\wb)}$.

To go from (\ref{csg}) to (\ref{wivg}) one performs the operation  $-(4 \pi)^{-1} \int d^2 w   V^{\wb} \partial^3_{\wb} $ on both sides of (\ref{csg}).  The LHS becomes the left term in (\ref{wivg}) (for a vector field with vanishing $\partial_w$ component).  The factors on the RHS are given by:
\be
-(4 \pi)^{-1} \int d^2 w   V^{\wb} \partial^3_{\wb} \J^i_{K^{+}_{(w,\wb)}}= -(4 \pi)^{-1} \J^i_W \label{intvp3k}
\ee
with
\be
W=  \int d^2 w   V^{\wb}  \partial^3_{\wb} K^{+}_{(w,\wb)}
\ee
(here we used the fact that $\J_V$ depends linearly on the argument $V$, so that the operation $\int d^2 w   V^{\wb} \partial^3_{\wb}$ can  be pushed inside the argument).
Finally, using 
\be
\partial^3_{\wb} \frac{(\zb-\wb)^2}{(z-w)} = - 4 \pi \delta^{(2)}(w-z) \label{D3K2}
\ee
one finds that $W= V^{\wb} \partial_{\wb}$. Thus one recovers the Ward identity for a vector field of the form $V^{\wb} \partial_{\wb}$. From the negative helicity CS theorem one can similarly obtain the Ward identity associated to a vector field of the form $V^{w}\partial_w$. By adding the resulting positive and negative helicity identities one obtains the general form (\ref{wivg}).

\subsection{External massive scalars}
Using the quantum version of (\ref{deltaVb})
\be
[b, \Qh_V] = i \L_{\Vh} b
\ee
the proposed Ward identity for external massive scalars takes the form:
\begin{multline}
- \frac{1}{4 \pi} \lim_{E_s \to 0^+} (1+ E_s \partial_{E_s})\\
 \hspace*{0.4in}\int d^2 w (   V^{\wb} \partial^3_{\wb} \bra {\rm out}| a^{\rm out}_+(E_s,w,\wb) \S  | {\rm in} \ket +  V^w \partial^3_w \bra {\rm out}|  a^{\rm out}_-(E_s,w,\wb) \S | {\rm in} \ket ) = \\  \sum_i \L_{\Vh^i} \bra {\rm out} |  \S | {\rm in} \ket.  \label{wardidVm}
\end{multline}
Here $\L_{\Vh^i}$ is the derivative $\Vh^\alpha \partial_\alpha$ on the $i$-th particle momentum variables (assumed to be outgoing; for incoming particles the factor is -$\L_{\Vh^i}$).

On the other hand the subleading soft theorem reads:
\be
\lim_{E_s \to 0^+} (1+ E_s \partial_{E_s})   \bra {\rm out} | a^{\rm out}_{+} (E_s,w,\wb) \S | {\rm in} \ket  = \sum_{i}  \frac{\e^+(w,\wb) \cdot p_i}{q \cdot p_i} \, \e^+_\mu(w,\wb) q_\nu  J^{\mu \nu}_i \bra {\rm out} |  \S | {\rm in} \ket . \label{csm}
\ee
with $J^{\mu \nu}_i$ the total angular momentum on the $i$-th massive particle.  We now repeat the steps that led to the equivalence in the massless case, starting in the direction soft theorem $\rightarrow$ Ward identity. Accordingly,  let us act with $-(4 \pi)^{-1} \int d^2 w   V^{\wb} \partial^3_{\wb} $ on both sides of (\ref{csm}) and with $-(4 \pi)^{-1} \int d^2 w   V^{w} \partial^3_{w}$ on the negative helicity soft theorem. Adding both terms we obtain an identity:
\begin{multline}
- \frac{1}{4 \pi} \lim_{E_s \to 0^+} (1+ E_s \partial_{E_s})\\
 \hspace*{0.4in}\int d^2 w (   V^{\wb} \partial^3_{\wb} \bra {\rm out}| a^{\rm out}_+(E_s,w,\wb) \S  | {\rm in} \ket +  V^w \partial^3_w \bra {\rm out}|  a^{\rm out}_-(E_s,w,\wb) \S | {\rm in} \ket ) = \\  \sum_i \L_{\Vt^i} \bra {\rm out} |  \S | {\rm in} \ket,  \label{wardidVmt}
\end{multline}
where
\be
\Vt^\alpha (\vec{p}/m) = \int d^2 w \, G^\alpha_B(\vec{p}/m; w,\wb)  \,V^B(w,\wb) , \label{Vt}
\ee
with
\be
 G^\alpha_{\wb}(\vec{p}/m; w,\wb) \partial_\alpha :=- \frac{1}{4\pi} \partial^3_{\wb} \frac{\e^+(w,\wb) \cdot p}{q \cdot p} \, \e^+_\mu(w,\wb) q_\nu  J^{\mu \nu}  , \label{greenV}
\ee
and similar expression for the $G^\alpha_{w}$ component. It can be verified this function takes the following simple form \cite{appendix}:
\be
G^\alpha_{B}(\vec{p}/m; w,\wb) \partial_\alpha  = -\frac{3 }{8 \pi} \frac{\sqrt{\gamma{(\qh)}}}{ \left( (q/E_s) \cdot (p/m)\right)^{4}}  (q^\mu/E_s)  \partial_B (q^\nu/E_s) J_{\mu \nu}.\label{greenV2}
\ee
Finally, parametrizing the 3-momentum particle as  $\vec{p}= m \rho \xh$ it can be shown $ G^\alpha_B(\rho,\xh; \qh)  \equiv G^\alpha_B(\vec{p}/m ; \qh)$ satisfies:
\be\label{august16-5}
(\Delta- 2) G^\alpha_B(\rho,\xh; \qh)=0 , \quad  D_\alpha G^\alpha_B(\rho,\xh; \qh)=0, \quad  \lim_{\rho \to \infty} G^A_B(\rho,\xh; \qh) = \delta^A_B \delta^{(2)}(\xh,\qh),
\ee
where $\Delta$ and $D_\alpha$ act on the $(\rho,\xh)$ coordinates.\footnote{The boundary condition in (\ref{august16-5}) can be obtained  by casting the $\rho \to \infty$ limit as a $m \to 0$ limit and using Eq. (\ref{D3K2}).}  It then follows that $G^\alpha_B$ is the Green's function for Eq. (\ref{eqxioalpha}), and so $\Vt=\Vh$.  Thus the identity (\ref{wardidVmt}) coincides with the Ward identity (\ref{wardidVm}). 

To go from (\ref{wardidVm}) to (\ref{csm}) we repeat the steps as in the massless case: Consider the Ward identity for the particular vector field
\be
V^A =  K^+_{(z_s,\zb_s)}.
\ee
By the same argument as in the pure gravity case,  the LHS  becomes the left term  in  (\ref{csm}). The differential operator on the  RHS is given by
\ba
\Vh^\alpha &= &  \int d^2 w \, G^\alpha_{\wb} (\vec{p}/m; w,\wb)  \,\frac{(\wb-\zb_s)^2}{(w-z_s)}  , \\
&=&  \frac{\e^+(w,\wb) \cdot p}{q \cdot p} \,  \e^+_\mu(w,\wb) q_\nu J^{\mu \nu} ,
\ea
where we used the definition (\ref{greenV}) of $G^\alpha_A$, integrated by parts and used Eq. (\ref{d3K}). One thus recovers the right term in (\ref{csm}). 

\section{Remarks}
From the results of the present paper together with those of the Maxwell case \cite{cl3} we arrive at  the general   picture that the `soft factors' that multiply the hard particles in the soft theorems have the role of `potentials' for the Green's functions that implement the large gauge transformation on such hard particles (either massless and massive; in the massless the ``Green's function'' is just an identity kernel).  We are referring here to identities  (\ref{D2soft}) and (\ref{greenst}) for the case of supertanslations and identities (\ref{D3K2}) and (\ref{greenV}) for the sphere diffeomorphisms (and similar identities for the Maxwell case \cite{cl3}). In each case, the 2-dimensional differential operator one is `inverting' are the ones arising in the soft part of the charge associated to the  large gauge transformation under consideration. It is so far unclear to us  what is the underlaying physical or geometrical mechanism behind such identities. We hope to clarify it in future investigations.

Another generic feature of the Green's functions and their potentials regards the relationship between the massive and massless cases: The identity kernel ``Green's function'' for the massless case can be obtained as a $m \to 0, \, \vec{p}= $ constant limit of the massive Green's function. 
 Mathematically this massless limit is equivalent to taking $\rho \to \infty$ keeping  $m=$constant.  From this perspective, the fact that the Green's function becomes an identity kernel ensures the correct boundary value of the gauge parameter that is being extended from null to time-like infinity.

Finally, we point out a subtlety regarding these potentials: Whereas the Green's functions are globally defined on the $\qh$ sphere, the potentials have a singularity at a point. The `potentials' we used have the singularity at the south pole $w,\wb=\infty$, which in turn is related to the fact that the expression for the polarization tensor used is only valid for soft graviton directions away from this point.   
When showing a soft theorem from a Ward identity (both massless and massive case),  an integration by parts is performed.  The smearing function being in integrated are such that the would be `boundary'  contribution  from the south pole vanishes. 

\section{Conclusions and outlook}
For the case of pure gravity, due to the seminal work of Strominger and his collaborators and many works which have followed, a clear picture is emerging as regards to the symmetry of quantum gravity S matrix in perturbative regime and its remarkable connections with some well known factorization theorems in the quantum gravity literature. However all the work done so far was in the context of pure gravity or at most in the context of pure gravity coupled to massless matter fields. This was due to the fact that the classical symmetry group of asymptotically flat geometries is most clearly understood at null infinity and is characterized in terms of so-called supertranslations and the Lorentz group (or its extensions like the local conformal symmetry or smooth diffeomorphisms of $S^{2}$) acting on the conformal sphere. 

A candidate symmetry group of semi-classical quantum gravity S matrix when the scattering states are massless is ${\cal G}$ and the associated Ward identities are equivalent to certain soft theorems. A natural question then emerges: Can we extend the above ideas to situations where perturbative gravity 
 couples to massive particles?  In this paper we argued that this generalized BMS group ${\cal G}$ can be made to act  not just on null infinity but also on time-like infinity. A key ingredient for such extension is a description of  time-like infinity as a constant curvature spacelike hyperboloid that  parametrizes  all unit time-like vectors  at infinity.
 
 
 We showed that there is a way to map generalized BMS vector fields and obtain a group of large diffeomorphisms which acts at the aforementioned notion of time-like infinity and proposed this group ${\cal G}$ as a candidate symmetry group of the S-matrix. We  showed that this group has a well defined action on the free states of the theory 
 and in fact the corresponding Ward identities are equivalent to Weinberg as well as Cachazo-Strominger soft theorems. As emphasized in the previous section, these equivalences rely upon intricate relationships between leading and sub-leading soft factors and Green functions which map supertranslation generators and sphere vector fields to functions and vector fields at the asymptotic hyperboloid  respectively.
 
A number of important issues remain to be understood. Here as in all previous papers we have worked with a formal S-matrix defined by assuming that asymptotic states are free states of the theory. However as has been well known for a long time, in the presence of long-range interactions asymptotic states of the theory are not free but are obtained by dressing free state with a coherent cloud of soft particles. This tension is already visible at the classical level. If we analyze the equation of motion of a massive scalar field coupled to linearized gravity, its solutions in the asymptotic $(\tau\rightarrow\infty)$ differ from the asymptotic limits of the free field by a phase factor  $\exp(i \frac{m}{2} h^{(1)}_{\tau\tau}\textrm{ln}\tau)$ where $h^{(1)}_{\tau\tau}$ is the coefficient of $\frac{1}{\tau}$  in a $\tau^{-1}$ expansion of the linearized metric.  The action of ${\cal G}$ vector fields will in general be ill-defined on such states. This is most clearly seen for the sphere vector fields which take the form $\xi^{\alpha}\ =\ \xio^{\alpha} + O(\tau^{-1})$. Their  action on dressed fields   contains a  multiplicative term proportional to $\ln( \t) \, \xi^\alpha \partial_\alpha  h^{(1)}_{\tau\tau}$. This implies that the action on the symplectic modes of massive fields, if we consider the fields to be dressed in contrast to free, is divergent. We believe that such divergences are intricately tied to the IR divergences that have been analyzed in \cite{akhoury} and these relations and their treatment remains an important question that remains unexplored.

\begin{appendix}

\section{Equations (\ref{eqxiot}) and (\ref{eqxioalpha})} \label{xiapp}

In this appendix we derive equations (\ref{eqxiot}) and (\ref{eqxioalpha}), which are the equations satisfied by the generators of ${\cal G}$ at time-like infinity. We first derive the differential equations in (\ref{eqxiot}),  (\ref{eqxioalpha}) and  then derive the boundary conditions. 
\subsection{Differential equations}
We start with the assumed expansion of the vector field at time-like infinity:
\ba
\xi^\t(\t,\rho,\xh) & = & \xio^\t (\rho,\xh) +O(\t^{-1}) \label{xitauapp}  \\
 \xi^\alpha(\t,\rho,\xh) & = & \xio^\alpha(\rho,\xh)+ \t^{-1} \xi^{(1) \alpha}(\rho,\xh)+ O(\t^{-2}) \label{xialphaapp}.
\ea
and compute  (\ref{boxxi}). In hyperbolic coordinates, the wave operator takes the form:
\be
\square=-\nabla^2_\t + \t^{-2} h^{\alpha \beta} \nabla_\alpha \nabla_\beta \label{boxhc}
\ee
where $h^{\alpha \beta}$ is the inverse of the metric in $\H$. The nonzero Christoffel symbols are:
\be
\Gamma^\t_{\alpha \beta}= \t h_{\alpha \beta}, \quad \Gamma^{\alpha}_{\beta \t} = \t^{-1} \delta^{\alpha}_{\beta},  \quad \Gamma^{\alpha}_{\beta \gamma},
\ee
with $\Gamma^{\alpha}_{\beta \gamma}$ the Christoffel symbols of the covariant derivative $D_\alpha$ on $\H$. Applying (\ref{boxhc}) to (\ref{xitauapp}), (\ref{xialphaapp}) one gets
\ba
\square \xi^\t &  = & \t^{-1} 2 D_\alpha \xio^\alpha + \t^{-2} \big( \Delta \xio^\t+ 2 D_\alpha \xi^{(1) \alpha}+3 \xio^\t \big)+ O(\t^{-3}) \label{boxxit}\\
\square \xi^\alpha & = & \t^{-2} \big( \Delta \xio^\alpha - 2 \xio^\alpha \big) +O(\t^{-2}). \label{boxxialpha}
\ea
The vanishing of the leading terms corresponds to the  first two equations in (\ref{eqxioalpha}). To obtain the equation for $\xio^\t$ we need further information. For this we now look at the divergence of the vector field:
\be
\psi:= \nabla_a \xi^a =  D_\alpha \xio^\alpha + \t^{-1}\big(3 \xio^\t+D_\alpha \xi^{(1) \alpha}\big)+O(\t^{-2}). \label{divxi}
\ee
Now, from equations (\ref{boxxi}) and (\ref{divzero}), it follows that  $\psi$ satisfies $\square \psi=0$ with fall-off $\psi=O(r^{-1})$ at null infinity. This is precisely the fall-off behavior of regular massless scalar fields. Such fields are known to decay at time-like infinity as   $\psi=O(\t^{-2})$, see for instance \cite{waldbook}.\footnote{We thank M. Reisenberger for help on this point.}  
The $O(\t^0)$ term in (\ref{divxi}) is already known to vanish from Eq. (\ref{boxxit}). The vanishing of the $O(\t^{-1})$ term gives us the missing information:
\be
D_\alpha \xi^{(1) \alpha} = -3 \xio^\t,
\ee
which combined with the vanishing of the  $O(\t^{-2})$ term in  (\ref{boxxit}) gives the first equation in (\ref{eqxiot}). 

\subsection{Boundary conditions}
We now write the vector field (\ref{xitau}), (\ref{xialpha}) in $(u,r)$ coordinates to impose the boundary condition (\ref{xibms}) at null infinity. The change of coordinates is given by:
\be
u   =  \t \big( \sqrt{1+\rho^2} - \rho \big) , \quad  \quad r = \rho \, \t .
\ee
We want to look at the $r \to \infty, u= \text{const}$ limit. In this limit, we have
\be
\rho \to \infty,  \quad \t \to \infty , \quad  \quad \t/(2 \rho) = \text{const.} ,
\ee
with 
\be
\t/(2 \rho) = u +O(r^{-1}).
\ee
Condition (\ref{xibms}) on the sphere components tell us:
\be
\xio^A = V^A +O(r^{-1}). \label{limxiA}
\ee
Writing the condition $D_\alpha \xio^\alpha=0$ together with (\ref{limxiA}) fixes the asymptotic form of $\xio^\rho$ to:
\be
\xio^\rho = - \rho \alpha +O(1), \label{xirho}
\ee
where $2 \alpha$ is the 2-d divergence of $V^A$. 
The $r$ component of the vector field has the asymptotic form:
\be
\xi^r = (\partial_\rho r) \xi^\rho +(\partial_\t r) \xi^\t = \tau \xio^\rho + ( \xi^{(1) \rho} + \rho \xio^{\t} ) +O(\t^{-1}),
\ee
and has to satisfy the condition (\ref{xibms}):
\be
\xi^r = - \alpha r +O(1).
\ee
From (\ref{xirho}) the  leading $- \alpha r$ term is recovered and we are left with the condition:
\be
\rho \xio^\t + \xi^{(1) \rho} = O(1). \label{taurho1}
\ee
We finally discuss the condition on the $u$ component.
Using 
\ba
\frac{\partial u}{\partial \t}  & = & \sqrt{1+ \rho^2}- \rho  = (2 \rho)^{-1}+ O(\rho^{-3}) ,\\
\frac{\partial u}{\partial \rho} & = & \t \left( \frac{\rho}{\sqrt{1+\rho^2}} -1\right) = - \t \big( (2 \rho^2)^{-1} + O(\rho^{-4}) \big),
\ea
one finds:
\be
\xi^u = - \t/(2 \rho^2) \xio^\rho + \xio^\t/(2 \rho) - \xi^{(1) \rho}/(2 \rho^2) + \ldots \label{xiu},
\ee
where the dots indicate subleading terms. This needs to  asymptote to:
\be
\lim_{r \to \infty} \xi^u = f + u \alpha . \label{limxiu}
\ee
From (\ref{xirho}) we see that the first term in (\ref{xiu}) reproduces the $u \alpha$ term in (\ref{limxiu}). We are then left with the condition:
\be
\lim_{\rho \to \infty}  \xio^\t/(2 \rho) - \xi^{(1) \rho}/(2 \rho^2) = f \label{taurho2}.
\ee
In order for (\ref{taurho1}) and (\ref{taurho2}) to be simultaneously satisfied, we must have:
\ba
\xio^\tau = \rho f +O(1) \label{limxitau} \\
\xi^{(1) \rho} = -\rho^2 f +O(\rho). \label{limxi1rho}
\ea
To summarize, the asymptotic condition (\ref{xibms}) translate into conditions (\ref{limxiA}), (\ref{xirho}), (\ref{limxitau}) and (\ref{limxi1rho}) for the vector field in $(\rho,\tau)$ coordinates.

\section{Equations (\ref{D2soft}) and (\ref{D2soft2})} \label{D2softapp} \label{appC}
The tensorial structure in $(w,\wb)$ implies $D^2_{\wb}$ acts as in  Eq. (\ref{D2itopz}). Performing this operation one finds:
\be
\partial_{\wb}\big( \gamma^{w \wb} \partial_{\wb} (\gamma_{w \wb} \s(z,\zb; w,\wb)) \big) = \frac{1 + w \wb}{1+z \zb}\big( 4 \pi \delta^{(2)}(w-z) + 2 \pi  (\wb -\zb) \partial_{\wb} \delta^{(2)}(w-z) \big), \label{D2softapp1}
\ee
where we used $\partial_{\wb}(w-z)^{-1}= 2 \pi \delta^{2}(w-z)$. Seen as a distribution we have that $(\wb -\zb) \partial_{\wb} \delta^{(2)}(w-z) = - \delta^{(2)}(w-z)$.  Furthermore by the Dirac deltas we can set $w=z$ in the prefactors of (\ref{D2softapp1}). It then follows that as a distribution:
\be
\partial_{\wb}\big( \gamma^{w \wb} \partial_{\wb} (\gamma_{w \wb} \s(z,\zb; w,\wb)) \big) = 2 \pi \delta^{(2)}(w-z)
\ee
which corresponds to Eq. (\ref{D2soft}).

We now show Eq. (\ref{D2soft2}). In the $(z,\zb)$ variables $\s(z,\zb; w,\wb)$ is a scalar. Hence the second derivative acts as: $D^2_{\zb}= \partial^2_{\zb} - \Gamma^{\zb}_{\zb \zb} \partial_{\zb} = \gamma^{z \zb} \partial_{\zb} \gamma_{z \zb} \partial_{\zb}$. Performing this operation one finds:
\be
\gamma^{z \zb} \partial_{\zb}( \gamma_{z \zb} \partial_{\zb} \s(z,\zb; w,\wb)) = \frac{1+w \wb}{1+ z \zb}\big( 4 \pi \delta^{(2)}(w-z)  -2 \pi  (\wb -\zb) \partial_{\zb} \delta^{(2)}(w-z) \big).
\ee
Identifying $(\wb -\zb) \partial_{\zb} \delta^{(2)}(w-z) = \delta^{(2)}(w-z)$ one finds
\be
\gamma^{z \zb} \partial_{\zb}( \gamma_{z \zb} \partial_{\zb} \s(z,\zb; w,\wb)) = 2 \pi  \delta^{(2)}(w-z) .
\ee
which corresponds to Eq. (\ref{D2soft2})

\end{appendix}

\end{document}